\input{aipcheck}
\documentclass[final,numberedheadings]{aipproc}
\usepackage{aas_macros}

\layoutstyle{8x11double}

\begin{document}

\title{Diffuse Extragalactic Background Radiation}


\author{Joel R. Primack}{
  address={University of California, Santa Cruz}
}

\author{Rudy C. Gilmore}{
  address={University of California, Santa Cruz}
}

\author{Rachel S. Somerville}{
  address={Space Telescope Science Institute, Baltimore, MD} 
}

\begin{abstract}
Attenuation of high--energy gamma rays by pair--production with UV, optical and IR background photons provides a link between the history of galaxy formation and high--energy astrophysics.  We present results from our latest semi-analytic models (SAMs), based upon a $\Lambda$CDM hierarchical structural formation scenario and employing all ingredients thought to be important to galaxy formation and evolution, as well as reprocessing of starlight by dust to mid- and far-IR wavelengths.  Our models also use results from recent hydrodynamic galaxy merger simulations.  These latest SAMs are successful in reproducing a large variety of observational constraints such as number counts, luminosity and mass functions, and color bimodality.  We have created 2 models that bracket the likely ranges of galaxy emissivities, and for each of these we show how the optical depth from pair--production is affected by redshift and gamma-ray energy.  We conclude with a discussion of the implications of our work, and how the burgeoning science of gamma-ray astronomy will continue to help constrain cosmology. 
\end{abstract}

\maketitle

\section{Introduction}
\begin{figure}
\resizebox{1.7\columnwidth}{!}{\includegraphics{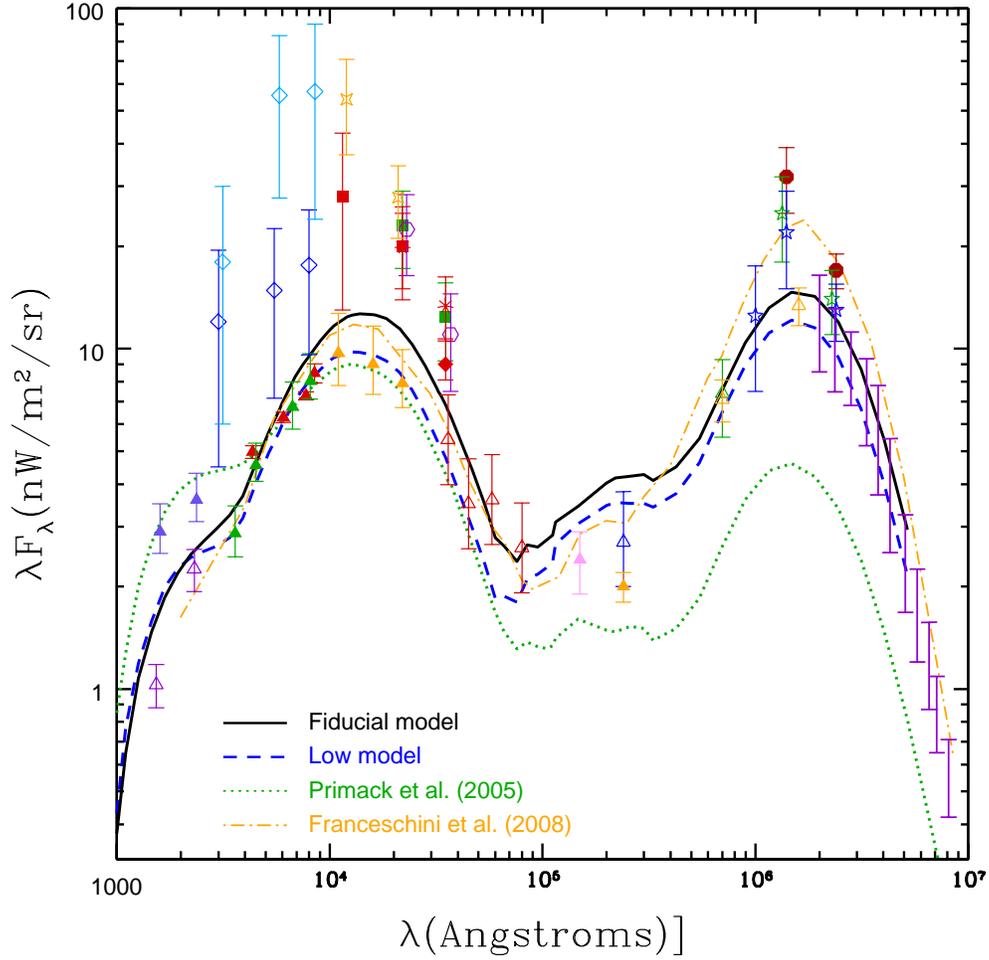}}
\caption{The predicted z=0 EBL spectrum from our fiducial (black) and low (dashed blue) models, compared with experiments at a number of wavelengths.  Our previous model \cite{primack05} is also shown for comparison (dotted curve).  We have also shown the model of Franceschini et al. \cite{franceschini08} for comparison (dash-dotted orange line).  Upward pointing arrows indicate lower bounds from number counts; other symbols are results from direct detection experiments.  Some points have been offset slightly to improve readability.  
{\bf Lower limits:}  The open blue-violet triangles are results from Hubble and STIS \citep{gardner00}, while the magenta triangles are from GALEX \cite{xu05}.  The green and red triangles from Hubble Deep Field \citep{madau00} and Ultra Deep Field \citep{dolch08} respectively, combined with ground based-data.  Gold triangles are also from this same work by Madau \& Pozzetti.  Open red triangles are from IRAC on Spitzer \citep{fazio04}, and the pink point at 15 $\mu$m is ISOCAM \citep{elbaz02} on ISO.  The remaining lower limits are from MIPS at 24, 70, and 160 $\mu$m on Spitzer \citep{papovich04,chary04,frayer06,dole06}.  {\bf Direct Detection:}  The higher open blue triangles are from Bernstein \citep{bernstein07}, while the lower points are the original determinations from \citep{bernstein02a}.  The high-reaching cyan points in the near-IR are from \citep{matsumoto05}. While the rest of the points in this region are based upon DIRBE data with foreground subtraction \citep{wright01} (dark red squares), \citep{cambresy01} (orange 4-stars), \citep{levenson08} (red diamond), \citep{gorjian00} (purple open hexes), \citep{wright&reese00} (green square), \citep{levenson07} (red asterisks).  In the far-IR, direct detection data is shown from DIRBE \citep{wright04} (blue stars), and \citep{hauser98} (green stars), and also purple bars showing the detection of FIRAS \citep{fixsen98}.}
\label{fig:eblflux}
\end{figure}

The extragalactic background light (EBL) consists of photons emitted at all epochs by stars and AGN, and subsequently modified by redshifting and dilution due to the expansion of the universe.  The bulk of the EBL energy is in two peaks in wavelengths, one in the optical and near--IR consisting of direct light from stars, and another in the far--IR producing by thermal radiation from dust (see Figure \ref{fig:eblflux}).  At still longer wavelengths, the background spectrum is dominated by the cosmic microwave background.  Emissions from polycyclic aromatic hydrocarbon (PAH) molecules also form an important component at a fixed series of wavelengths in the mid--IR.

Because the production of the EBL is directly linked to the star formation history of the universe, limits on the EBL can be used to provide constraints on the history of galaxy formation and evolution.  Measurement of this light through direct observation is complicated by foreground emission from our own galaxy and reflected zodiacal light from our sun, which are much brighter than the EBL across most of the optical and IR spectrum \citep{hauser&dwek01}.  Interplanetary dust is the major contributor of foreground light at most wavelengths, with starlight becoming substantial in the optical and near-IR, and the interstellar medium most important in the submillimeter regime.  In a series of papers by Rebecca Bernstein \citep{bernstein02a,bernstein02b,bernstein07} the optical EBL was estimated by measuring the total background in 3 HST WFPC2 bands, and subtracting away the zodiacal and galactic contributions.  The near-IR flux has been measured by the DIRBE instrument on the Cosmic Background Explorer \citep{wright&reese00,wright01,gorjian00,cambresy01,levenson07}, using foreground source subtraction techniques and modeling of the zodiacal light, and yielding high estimates in this range.  Results from IRTS \citep{matsumoto05} claimed an even higher background level below 2 $\mu$m a level which would require a large contribution from an as-of-yet undetected source type.  An analysis of the fluctuations observed by DIRBE \citep{kashlinsky00} yielded upper limits to the EBL in the mid to far-IR, but this author's interpretation of a high background arising from sources at $z \geq 8$ has been challenged by the results of other studies \cite{thompson08,cooray07}.  In the far-IR regime, observations from the COBE FIRAS \citep{fixsen98} and DIRBE \citep{hauser98} instruments provide direct measurements of the background radiation.  At shorter wavelengths (60 and 100 $\mu$m) the DIRBE instrument places upper limits on the background.  Analyses of the DIRBE data at far-IR wavelengths have been attempted by other authors using new models for the zodiacal light and instrument calibration \citep{finkbeiner00,lagache00,wright04}.  The rather high detection claimed by Finkbeiner et al. at 60 and 100 $\mu$m, which required careful subtraction of the bright foreground, has been disputed by other authors \citep{puget&lagache01}.  

Integration of galaxy counts is a way to set firm lower limits on the EBL, although the degree to which these measurements converge on the true value remains controversial.  Measured luminosities of galaxies in the Hubble Deep Field (HDF) \citep{madau00} place a lower bound on EBL fluxes in the 0.4-1 $\mu$m range, a much lower level than that proposed in the Bernstein papers, with 2MASS data extending counts in this paper to 3 $\mu$m.  It is argued in this work that the flatness of the faint counts in all bands (flatter than 0.4 on a log-counts vs. magnitude diagram) indicate convergence, and therefore a low EBL across the optical and near-IR.  A similar analysis conducted using IR data from the Subaru deep field \citep{totani01} modeled possible selection effects and supported the conclusion of a low optical--near-IR background compared to direct detection claims.  In the UV, limits exist from GALEX \citep{xu05} and observations of the HDF with the STIS instrument \citep{gardner00}, with the latter finding a considerably higher bound.  The IRAC instrument on {\it Spitzer} has placed lower limits on several bands in the near-- to mid--IR \citep{fazio04}, which similarly are well below the IRTS and DIRBE direct detection fluxes.  In a complementary approach to their DIRBE sky-subtraction papers, Levenson and Wright used IRAC data to calculated the best-fit flux at 3.6 $\mu$m using a profile-fit to estimate the light from the unobservable faint fringes of galaxies, and a broken power law model for the number count distribution \citep{levenson08}. ISOCAM and the MIPS instrument on Spitzer have reported lower limits from number counts at 15, 24, 70, and 160 $\mu$m~ \citep{elbaz02,chary04,frayer06,papovich04,dole06}.  Galaxy counts from the SCUBA instrument provide a lower limit at 850 microns \citep{coppin06}, with an estimated 20 to 30\% of the background at these wavelengths resolved into point sources.  Our previous EBL model \cite{primack05} was systematically low in the mid- and far-IR, and we warned in that publication that the model should not be trusted in this regime.  Our new models, presented here, do a much better job in matching constraints from number counts.

\begin{figure}
\resizebox{1.0\columnwidth}{!}{\includegraphics{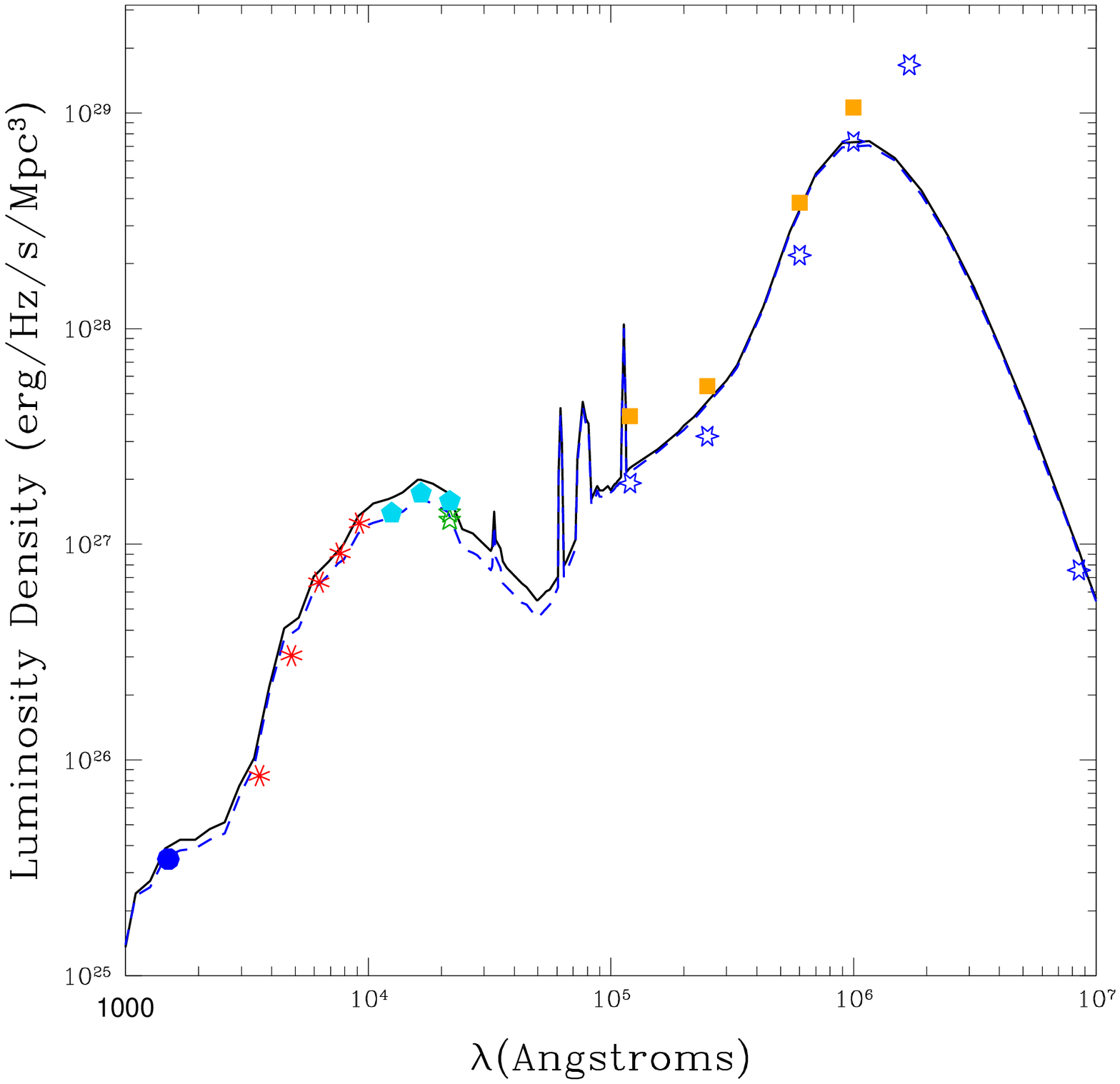}}
\caption{The z=0 luminosity density of our model.  As before, the fiducial model is shown as a solid black line, the low model as dashed blue.  Data at a number of wavelengths is shown from GALEX, SDSS, 6dF \citep{jones06}, 2MASS \citep{cole01,bell03}, and IRAS \citep{takeuchi01}.}
\label{fig:lumdens_z0}
\end{figure}

The EBL provides an important test of star-formation observations; if there is a large gap between the flux from sources in deep surveys and direct measurements, then an unresolved population must exist to provide the missing energy.  The present day EBL is linked to the history of star formation through the stellar initial mass function (IMF), as well as the evolving metallicity and dust distribution, as we will discuss later.  It is well established that star formation rates per unit cosmological volume were much higher in the past, peaking at roughly z$\sim$2 \citep{hopkins04,gabasch04}.  

Studies of the buildup of stellar mass have typically found that the integrated star formation rate tends to exceed stellar mass as measured by IR (fossil) light from old stars, with the discrepancy becoming worse with redshift.  The local star-formation rate is typically measured from the strength of the H$\alpha$ (Balmer) line.  At higher redshift, other tracers must be used.  These can include UV continuum emission corrected for dust, radio emission by supernovae remnants, emission lines which are excited by UV radiation from young stars, or measuring reprocessed light from dust in the infrared.  All of these methods are fraught with uncertain amounts of contamination from quasars and biases from dust and metallicity, and different methods can produce widely varying results for the same galaxy.  An overview of these issues can be found in \citep{hopkins&beacom06}.  One solution to the SFR--fossil mass discrepancy may lie in a variable IMF which suppresses formation of low mass stars in starbursts occurring primarily at higher redshift \citep{fardal07,dave08}.  In this way, high mass stars that produce tracers of the star-formation rate could be produced at a high rate while lower mass stars that account for most of the integrated stellar mass are created in smaller quantities.  Alternatively, it is possible that systematic biases in determinations of the star formation rate at increasing redshift are responsible for the discrepancy.  This can be phrased in terms of measurements of Dav\'{e}'s star-formation activity parameter, which determines the Hubble times required for galaxy to reach their current mass at current SFR, and is seen to decrease too much between present day and z$\sim$2.  This conflicts with model predictions, and the passive population that would be required to counterbalance the high star formation in these rapidly growing systems is not observed.  The recent work of \citep{chen08} has found lower specific star formation rates in galaxies from the SDSS and DEEP2 surveys than typically measured by more usual means using an alternative method based on higher order Balmer lines.  The lower normalization found in this model can be attributed to a number of factors, though the possibility of a changing IMF cannot be ruled out.

Unresolved primordial (population III) stars could in principle provide a large contribution to the background light, and the near-IR peak observed by the IRTS satellite \citep{matsumoto05} has been interpreted as the redshifted photons from massive population-III stars beyond redshift $\sim$9 \citep{salvaterra&ferrara03}.   This peak reaches nearly a factor of 10 higher than levels from resolved number counts, but this claim is at odds with \citep{madau00,totani01,thompson08,cooray07} in which it is argued that the contribution from resolved galaxies has nearly converged, and there is little room for any additional sources such as an early generation of stars with a top-heavy IMF.  The levels of metal formation from massive stars in these models required metals to either be locked away in compact products or dispersed in the IGM in a very inhomogeneous fashion.  The large fraction of baryons which would have to be processed through primordial stars, $\sim$10\%, and the lack of J-band dropout detections of these sources have strongly disfavored this interpretation \citep{salvaterra&ferrara06,dwek05}, as does amount of material that would be contained in intermediate mass black holes \cite{madau&silk05}.  Additionally, an extragalactic origin for this feature would have a huge impact on the optical depth of TeV gamma rays, as we will discuss shortly. 

Three approaches have been followed to calculate the EBL (as described
by \cite{kneiske02}): (A) Evolution Inferred from
Observations -- e.g., [Kneiske et al. 2002, Franceschini et al. 2008];
(B) Backward Evolution, which starts with the existing galaxy
population and evolves it backward in time -- e.g., \cite{steckermalkan&scully06}; and (C) Forward Evolution, which begins with
cosmological initial conditions and models gas cooling, formation of
galaxies including stars and AGN, feedback from these phenomena, and
light absorption and re-emission by dust -- e.g., \citep{primack01,primack05}.  We will concentrate on Forward
Evolution here, based on semi-analytic models (SAMs).

\section{Semi-Analytic Modeling}

\begin{figure}
\resizebox{1.0\columnwidth}{!}{\includegraphics{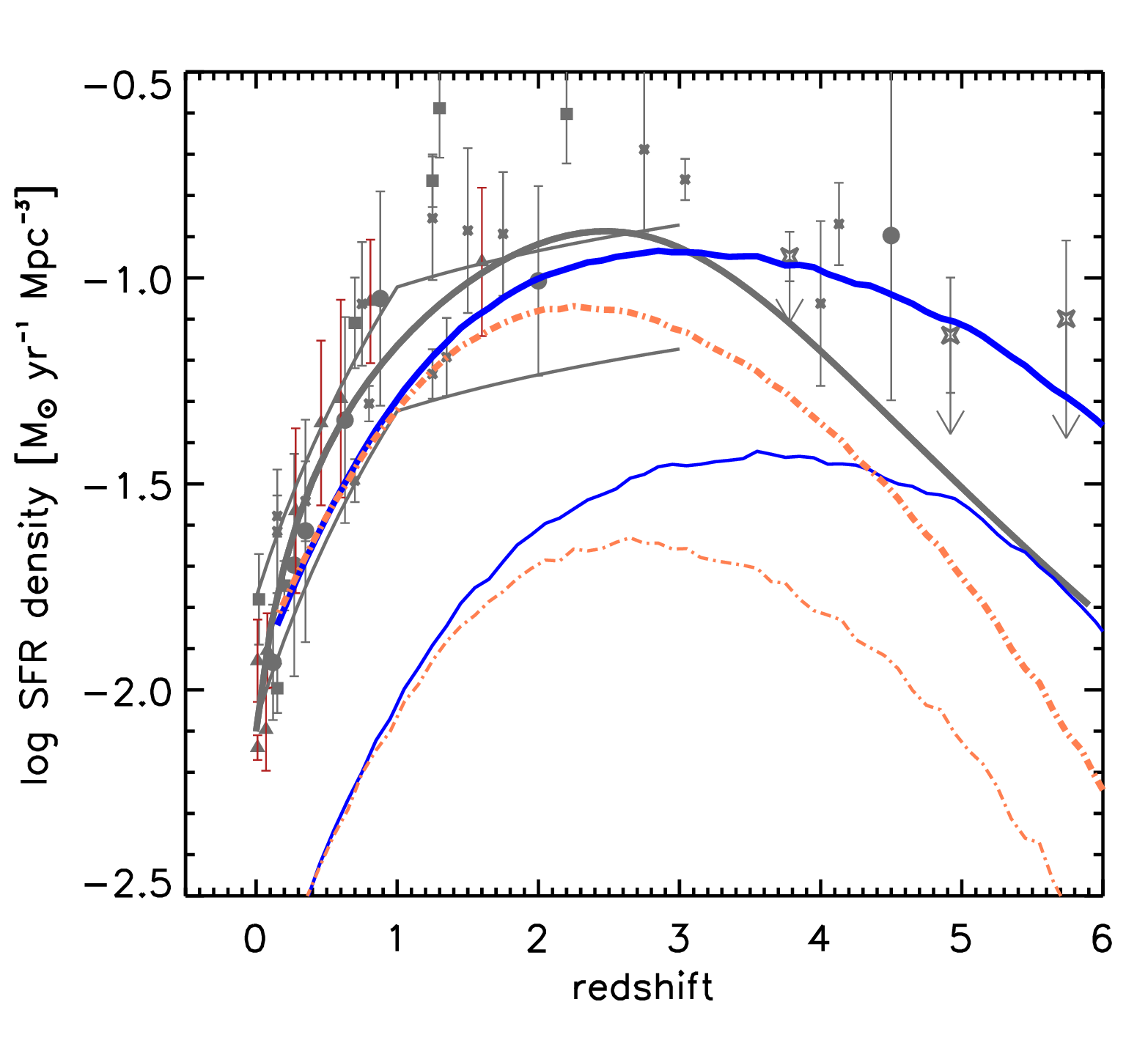}}
\caption{The star formation rate density for the fiducial (blue) and low (broken orange) models; this plot is from \cite{somerville08} (Figure 14).  Lower curves of the same colors indicate the contribution from starbursts.  The data in this plot is from the compilation of \cite{hopkins04} and has been converted to a Chabrier initial mass function.  The gray line is the best fit to the data from \citep{hopkins&beacom06}.  The thin gray broken contours are estimates from GALEX observations \cite{schiminovich05}}
\label{fig:sfr}
\end{figure}

\begin{figure}
\resizebox{1.0\columnwidth}{!}{\includegraphics{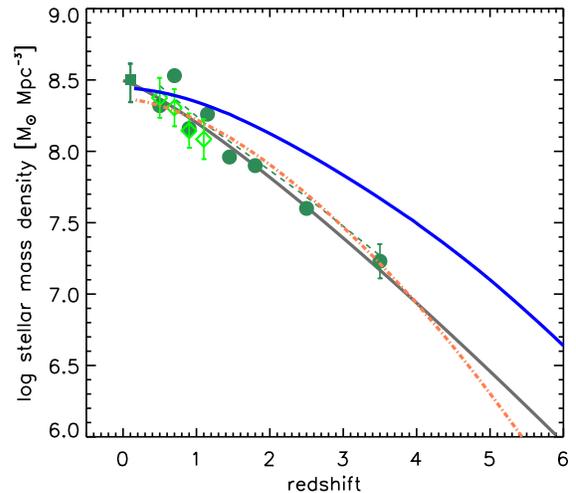}}
\caption{The integrated stellar mass vs redshift in each of our models, also from \cite{somerville08} (Figure 14).  As in the previous figure, the solid blue curve denotes the fiducial model, and broken orange the low model.  The solid square is the z=0 estimate of \citep{bell03}, the circles are from \citep{fontana06}, and the open diamonds are from the COMBO-17 estimates of \citep{borch06}.  The gray line is the best fit to the observational compilation of \citep{wilkins08}.}
\label{fig:mstellar}
\end{figure}

\begin{figure}
\resizebox{2.0\columnwidth}{!}{\includegraphics{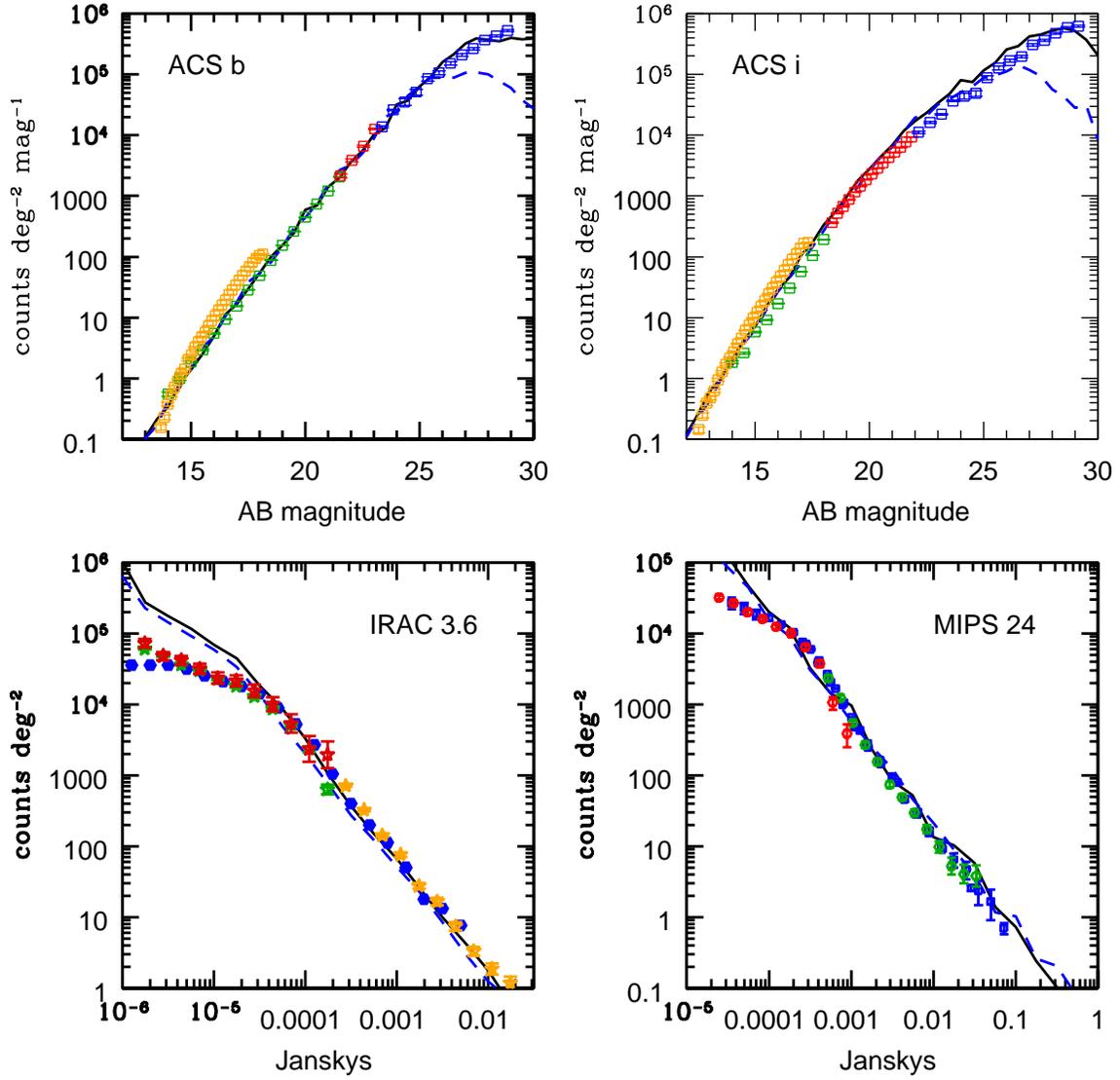}}
\caption{Number counts in our model at a variety of wavelengths.  The top two panels show in two of the Hubble ACS bands; as in previous figures the solid black line represents the fiducial model, and the dashed blue line the low model.  Red, blue and green data is from the compilation by \citep{dolch08}, which includes data from the Hubble Ultra-Deep Field.  Additional data from SDSS-DR6 is provided by \citep{md&prada08}.  The lower two plots are infrared bands (Spitzer IRAC and MIPS).  Data in the IRAC 3.6 $\mu$m band is from \citep{fazio04,sanders07}; the MIPS data is from \citep{papovich04,shupe08,chary04}.}
\label{fig:nc}
\end{figure}


The use of semi-analytic models to study galaxy formation was pioneered by \citep{white91}, and initial work was carried out primarily by two groups based in Durham and Munich.  Our group was the first to develop a complete SAM code for galaxy formation \citep{somerville&primack99} and use it to model the EBL.  Our code used a merger-tree formalism which was consistent with high-resolution dark matter simulations of the time, and several models of star formation and supernovae feedback were investigated.  In \citep{somerville01}, these models were combined with the improved dust emission code of \citep{devriendt99,devriendt00} for a prediction of the EBL and absorption of gamma rays \citep{primack01}.  At the 2004 Symposium on High Energy Gamma Ray Astronomy our group presented a modified spectrum and attenuation predictions \citep{primack05}.  This SED was somewhat lower than our 2001 result \cite{primack01}, primarily because our 2005 model \cite{primack05} had been recalibrated to fit the local luminosity density as determined by new surveys such as 2MASS \citep{cole01}, SDSS \citep{blanton03}, and 2dF \citep{norberg02}.  While the prediction of this model was consistent with optical and near-IR number counts data, it was below both the ISOCAM lower limit at 15 microns and the direct detection fluxes of DIRBE and FIRAS in the far-IR, indicating that the light remission by dust was being significantly underrepresented.      

Our current SAM \cite{somerville08} is normalized using a variety of updated survey data, particularly in the IR, and also recent measurements of the fundamental cosmological parameters.  In Figures \ref{fig:eblflux} and \ref{fig:lumdens_z0} we show the present--day extragalactic background light and luminosity density for two models of galaxy evolution.  Our `fiducial' model is based upon a concordance cosmology with $\Omega_m$ = 0.3, $\Omega_{\Lambda}$ = 0.7, $H_0$ = 70.0, and $\sigma_8$ = 0.9.  Our `low' model is based upon the best fit values from WMAP3, which include a lower $\sigma_8$ of 0.761.  This lower normalization of the primordial power spectrum leads to delayed structure formation and lower EBL.  The most recent data from WMAP5 \citep{komatsu08} suggests an intermediate value of $\sigma_8$ = 0.817$\pm$0.026, so we believe these two models can serve as bracketing the reasonable possibilities arising from cosmological uncertainty.  The SAMs used here are based either upon the merger history of the dark matter halos in the simulation or else the extended Press-Schechter method, in which for each present-day halo, a merger 'tree' is constructed, using the method similar to that described in \citep{somerville&kolatt99}.  The merger trees track the buildup of dark matter mass, with junctions representing halo mergers which form larger, virialized halos.  The NFW profile \citep{navarro97} is used as the initial halo description with the concentration determined using a fitting formula based on \citep{bullock01}.  The model does not account for scatter or merger history in determining concentration.  These methods do allow tidal disruption and destruction of halos in minor mergers to be considered; if a halo is destroyed prior to merging, its stars join the diffuse stellar component around the central galaxy.

Gas which cools around the halo is assumed to initially fall into a thin disk with exponential profile.  The scale radius of this disk is determined using conservation of angular momentum and the concentration and baryon fraction of the disk.  Our model computes the cooling time for gas based upon density, metallicity, and the initial virial temperature.  Our recipe agrees well with gas infall and cooling rates from 3-dimensional hydrodynamic simulations of cold and hot flows \cite{birnboim&dekel03,dekel&birnboim08,keres08}. 

Feedback from supernovae can heat the cold gas reservoir and drive it from the galaxy.  This gas will either be deposited in the hot reservoir of the galaxy, or returned to the IGM, depending on the wind velocity relative to the virial velocity of the halo.  Ejected gas can cool and return to the galaxy on a timescale roughly equal to the dynamical time of the galactic halo.  The model discriminates between cold- and hot- mode accretion of gas based on the relative values of the cooling time of the gas vs dynamical time of the halo.    

Star-formation in our model occurs in two regimes, quiescent star-formation and merger-driven starbursts.  The contribution from each of these modes can be seen in Figure \ref{fig:sfr}.  The integrated mass vs. redshift is plotted in Figure \ref{fig:mstellar}.  The quiescent star formation is treated using a recipe based on the Schmidt-Kennicutt law \citep{kennicutt89,kennicutt98} with slope 1.4 and normalization to the Chabrier IMF.  Star formation is assumed to cut off below a specific surface density, giving the exponential galactic disks a corresponding radius within which stars are forming.   Merger-driven bursts are parametrized by the mass ratio of the merging pair, the mass in this case being the total mass in the inner part of the halo, taken to be twice the characteristic NFW scale radius.  The burst efficiency parameter determines the fraction of cold gas converted to stars in the burst, and no burst occurs for a mass ratio of less than 1 to 10.  The functional form taken for this parameter is from \citep{cox08}.  The SFR during the burst is proportional to the available fuel, and therefore takes the form of a decaying exponential after an event.  Details of the functional forms for the burst efficiency and timescale can be found in \citep{somerville08}. 

Our implementation of AGN feedback is similar in many respects to that of \citep{sijacki07}, in which AGN operate in `bright' and `radio' modes, with the latter being switched on when the accretion rate rises above a critical value.  All bright mode accretion is triggered by galaxy mergers, and the interdependent processes of AGN activity and black hole growth are based upon results from a large suite of hydrodynamic simulations.  Every top-level halo in our simulation begins with a seed black hole of 100 solar masses; we do not find results to be sensitive to this amount.  Black holes in merging galaxies coalesce rapidly and the product grows at the Eddington rate until reaching a critical `blowout' mass, after which accretion rate falls as a power law, until it is cut off completely upon attaining a final mass determined by the spheroid mass and gas fraction.   Radiative momentum from the accreting black hole is transferred to the galactic wind with an assumed coupling efficiency.  In radio mode, the black hole enters a phase of Bondi-Hoyle accretion \citep{bondi52}.  The net cooling in this mode is the cooling rate minus the heating from Bondi accretion.  Heating is ignored if the cooling time is shorter than the dynamical time of the halo (`cold mode' cooling).

Predictions for galaxy number counts in the optical and IR are shown in Figure \ref{fig:nc}.  The detection of the EBL in the far-IR by DIRBE and FIRAS at a level comparable to the direct emission from starlight requires the inclusion of galaxy populations very different from the Milky Way and other nearby galaxies, which emit most light in the optical and near-IR.  Detectors in the mid-IR such as ISOCAM and MIPS find significant IR emission from star-forming galaxies.  The cause of this is primarily tiny particles of dust released in supernovae explosions, with typical sizes ranging from nanometers to tenths of a micron.  The temperature of these grains is only weakly affected by the flux of the surrounding radiation field, due to the blackbody emission which increases as the 4th power of temperature and the falloff of emission at long wavelengths, and therefore the peak emission ranges from about about 170 microns for a Milky Way type spiral to 60 microns for a ULIRG \cite{lagache05}.  Poly-aromatic hydrocarbons (PAHs) are another class of absorbers; tiny particles which emit at a group of specific wavelengths in the mid-IR from 3 to 17 microns.  The absorption and re-emission of light, primarily from the UV, by dust and PAH particles in our model is accounted for using the \textsc{stardust} templates of \citep{devriendt99,devriendt00,guiderdoni99}, which are embedded in our semi-analytic model, and account for emission at wavelengths from the far-UV to sub-millimeter.  These models are based upon the IRAS color-luminosity relation, and take the age, star-formation rate timescale, and the size of the gaseous disk as inputs.  The metallicity is tracked both in the stellar spectra and surrounding gas, with predictions for radiative transfer based on geometry and metal distribution modeled after the Milky Way and nearby galaxies.  We are working on improved dust models that will be more self-consistent.

\section{Gamma-Ray Attenuation}

\begin{figure}
\resizebox{1.0\columnwidth}{!}{\includegraphics{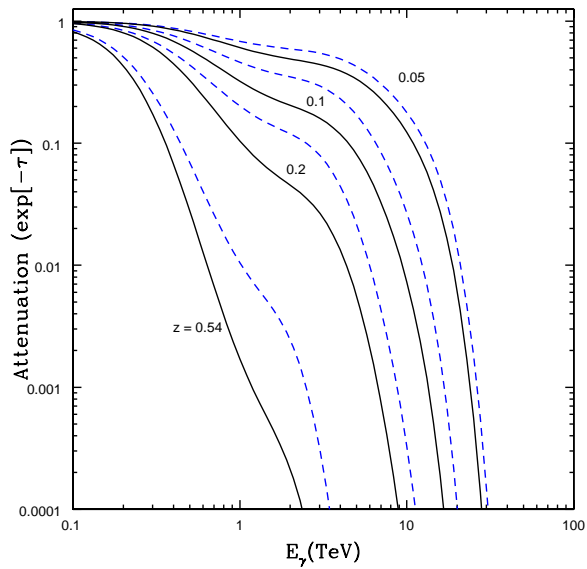}}
\caption{The attenuation $e^{-\tau}$ of gamma-rays vs. gamma-ray energy, for sources at $z= 0.05$, 0.1, 0.2, and 0.536 (the redshift of 3C279).  As previously, the fiducial model is solid black and the low model is dashed blue.  Increasing distance causes absorption features to increase in magnitude and appear at lower energies.  The plateau seen between 1 and 10 TeV at low redshift is a product of the mid-IR valley in the EBL spectrum.}
\label{fig:opdep}
\end{figure}

A consequence of the EBL is the attenuation of high--energy gamma rays through electron--positron pair production \citep{gould67}.  The threshold for this process occurs when the center of mass energy of the photons is equal to the rest mass of the electron-positron pair.  The cross section is maximized at approximately for center of mass energies of approximately twice the threshold, and falls as inverse energy for $E \gg E_{th}$.  Gamma rays above 1 TeV are most attenuated by the near- and mid-IR range of the EBL. Those in the 300 GeV to 1 TeV regime are sensitive to light in the near-IR and optical; below 300 GeV only UV photons have sufficient energy to cause the pair-production interaction.  Below 20 GeV there is little absorption due to the increasing scarcity of hard UV background photons \cite{gilmore08}.

Observations of the attenuation of gamma-ray spectra from extragalactic sources provide a measurement of the EBL.  We predict this attenuation here based on propagation of gamma-rays through the evolving EBL given by our fiducial and low models.  In Figure \ref{fig:opdep}, we show the optical depth $\tau$ (attenuation $=e^{-\tau}$) for gamma-rays vs. energy for a variety of redshifts. In principle, the cosmological history of the EBL could be reconstructed by comparing observations of high-energy sources at different redshifts to their known intrinsic spectra.  Unfortunately, the emission mechanisms of GeV and TeV sources are in general poorly understood.  Ground-based gamma-ray astronomy, which began largely with the Whipple and Hegra collaborations in the 1990s, has opened up the energy range above $\sim 100$ GeV.  Today exploration in this regime is led by the VERITAS \citep{maier07}, H.E.S.S. \citep{hinton04}, and MAGIC \citep{cortina05} experiments.  The recently-launched Fermi (GLAST) and AGILE satellites \citep{tavani08} will provide much needed data.  The LAT instrument on Fermi has an energy range of 20 MeV to 300 GeV, and will efficiently scan the sky in this energy range.  In addition to detecting sources on its own, Fermi will act as a finder for pointed observations by ground based telescopes.  While in survey mode, the LAT will see the entire sky every 3 hours, and will be able to report flaring blazars to other detectors for study.

\begin{figure}
\resizebox{1.0\columnwidth}{!}{\includegraphics{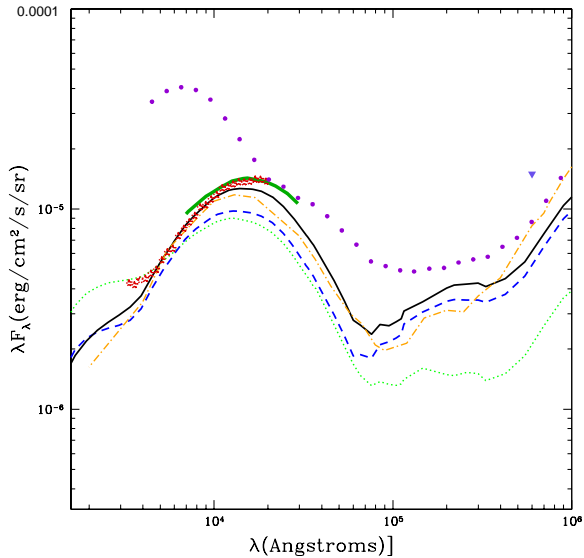}}
\caption{Here we show limits from gamma-ray experiments compared to the 4 models presented in Figure \ref{fig:eblflux}.  These limits are based upon the $\Gamma \geq 1.5$ criterion discussed in the text.  The thick green curve is the limit from \cite{aharonian06} and the red stars are from the observations of the 3C279 by MAGIC \cite{albert08}.  The purple points are from functional EBL forms of \citep{mazin&raue07} and the point at 60 $\mu$m is from \citep{dwek&krennrich05}.}
\label{fig:eblflux_gamlims}
\end{figure}

Ground-based detectors searching above 100 GeV have identified 23 extragalactic sources at the time of writing, including 21 BL Lac objects, radio galaxy M87, and the flat-spectrum radio quasar 3C279.  With the exception of M87 these objects are all blazars, accreting AGN which generate tightly-beamed relativistic jets that lie at a small angle relative to our line of sight.  Experiments at a wide variety of energy ranges have revealed that the spectra of blazars consist of a two-peaked distribution, the lower peak between the infrared and x-rays, the upper in the gamma-rays.  This pattern is commonly understood as the result of a relativistic jet of charged leptons.  The lower peak is attributed to synchrotron emission, while the gamma-ray peak is the result of inverse Compton radiation seeded by either the synchrotron photons themselves (the synchrotron-self Compton or SSC model) or photons from other sources (the external Compton or EC model).  It has been argued in the case of at least 2 rapidly varying blazars that such variability is better modeled with an external radiation source \citep{begelman08}.   There is evidence that there exists a BL Lac sequence, with less luminous objects having both peaks at higher energies than more brilliant sources \citep{fossati98,ghisellini98,maraschi08}, the extreme case being the HBL sources, in contrast to the brighter LBL sources.  The sequence may arise as a result of increased cooling of the accelerated particles at higher luminosities, suggesting that sources with brighter bolometric luminosities would have lower Lorentz factors and peak energies \citep{ghisellini98, ghisellini02}.  While they account for almost all the detected sources above 100 GeV, BL Lac objects are themselves only a small subset ($\sim$20$\%$) of all blazar sources, the other 80 percent being flat spectrum radio quasars like 3C279. 

\begin{figure}
\resizebox{1.5\columnwidth}{!}{\includegraphics{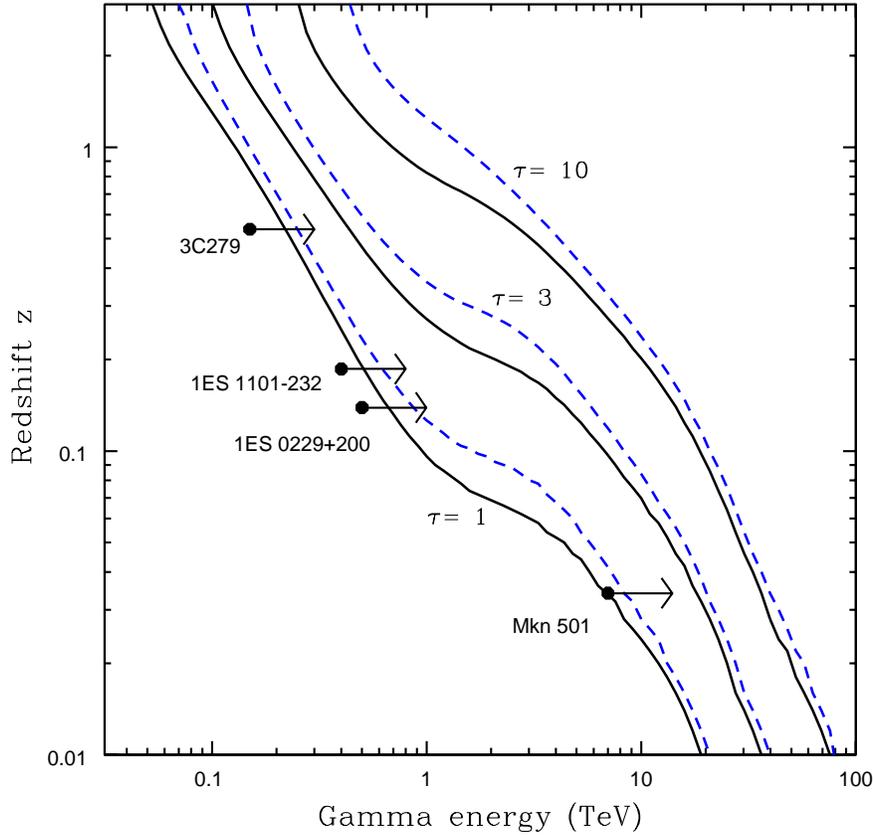}}
\caption{The gamma ray attenuation edges for the fiducial (black) and low (dashed blue) models.  The curves show the redshift at which the pair-production optical depth $\tau$ reaches the indicated value for a particular observed gamma ray energy.  The pairs of curves from lower left to upper right are the contours for $\tau=$ 1, 3, and 10.  The limits set by several observed sources are shown.}
\label{fig:attedge}
\end{figure}

While uncertainties and likely variation in the intrinsic spectrum of blazars make it difficult to link the observed spectra directly to EBL attenuation, it is possible to translate limits on the spectra to EBL constraints.  The standard assumption in placing limits the EBL from individual spectra is that the reconstructed intrinsic spectrum should not have a spectral index harder than 1.5; that is, $\Gamma \geq 1.5$ if $dN/dE \propto E^{-\Gamma}$ for photon count $N$, or alternatively $dF/dE \propto E^{-(\Gamma-1)}$ for flux $F$.  The standard value for a single-zone SSC spectrum is $\Gamma=(\alpha+1)/2$, where $\alpha$ is the spectral index of the shock accelerated electrons, and is not harder than 1.5 in most models \citep{aharonian01}.  This can be invalidated by assuming a non-standard spectrum for the electrons; a low energy cutoff in the electron energy will lead to inverse-Compton accelerated photons with an index as low as $\Gamma=2/3$ \citep{katarzynski06}.  In Figure \ref{fig:eblflux_gamlims} we show a number of constraints on present-day EBL experiments from gamma-ray observations.  The most recent limits on the EBL come from observations of blazars at more distant redshifts (z$>$0.1) that have been detected by the current generation of ground-based ACTs.  Observation by H.E.S.S. of two blazars at z=0.165 and 0.186 were used to set limits on the near-IR EBL based on the $\Gamma \geq 1.5$ criterion \citep{aharonian06}; in this case the maximal limit was our 2001 model \citep{primack01} multiplied by a factor of 0.45.  It should be noted that our 2005 EBL prediction \citep{primack05} included an optical and near-IR flux even lower than this older model with the multiplier of 0.45.  Another paper by the H.E.S.S. group set constraints from blazar 1ES 0229+200 at z=0.1396 \citep{aharonian07}.  While this blazar is a closer source than the two featured in the 2006 publication, the observed spectrum extended above 10 TeV and therefore probed the background in the mid-IR.  In this regime, the effect of optical depth on spectral form is minimal due to the approximate $\lambda^{-1}$ falloff in EBL flux.  The limits derived in this case are well above our 2005 model for a couple of different spectral slopes considered.  The most distant source observed at very high energies at the time of writing is quasar 3C279 at z=0.536, observed by the MAGIC experiment during a flare in February 2006 \citep{teshima07}.  The spectrum observed was quite steep, $4.1\pm0.7_{stat}\pm0.2_{sys}$, and extended from about 80 to nearly 500 GeV.  An analysis of the spectral modification \citep{albert08} found that there was little room for an EBL flux in the optical higher than one consistent with lower limits from number counts, approximately equivalent to the model of \citep{primack05} and with our fiducial and low models presented here -- see Figure \ref{fig:attedge}.  An alternative analysis of the spectral deconvolution of 3C279 by \citep{stecker08} disputed this analysis and argued that the higher EBL of \citep{steckermalkan&scully06,steckermalkan&scully07} could still lead to a steep best-fit spectrum.  Another approach to the problem is to attempt to constrain the EBL by using spectra from several sources simultaneously.  Dwek and Krennrich~\citep{dwek&krennrich05} considered 12 such permutations, and derived an upper limit at 60$\mu$m by declaring invalid those realizations leading to unphysical intrinsic blazar spectra with sharply rising TeV emissions.  An effort by \citep{costamante03} used observations of 4 blazars by HEGRA to limit the EBL in the near-IR and at 60 microns with limiting $\Gamma \geq$ 1.0 and 1.5.  More recently, this method was used in \citep{mazin&raue07}, who applied constraints from all observed TeV blazars to a large number of possible EBL functional forms created using a spline interpolation across a grid in flux versus wavelength space.  The lower bound of the union of excluded models formed an envelope representing the highest possible background that does not violate any constraints.  This was done for `realistic' and `extreme' bounds of $\Gamma \geq 1.5$ and $2/3$ respectively, and provided a limit on the EBL from the optical to the far-IR.


\section{Discussion}

Figure \ref{fig:attedge} shows the gamma-ray attenuation edge plot for our models.  This is a complementary plot to the optical depths of Figure \ref{fig:opdep}, and shows the distance to which gamma-ray detectors can view before encountering various levels of obscuration from photon-photon interactions. 

New self-consistent hybrid SAMs based on physical scaling from
numerical simulations and calibrated against empirical constraints now
enable us to interpret the relationship between galaxies, supermassive
black holes, and active galactic nuclei across cosmic history.  The
latest semi-analytic models are greatly improved over the best SAMs
that were available four years ago, largely because the improving
observations of galaxies challenged theorists to do better.  Although
the $\Lambda$CDM dark matter backbone for structure formation was
already well understood, earlier $\Lambda$CDM-based galaxy formation
models suffered from a set of interlinked problems -- especially
overcooling leading to excessive star formation in galaxies and
clusters, and a resulting failure to produce the observed red/blue
galaxy color bimodality. New SAMs, including the one \cite{somerville08}
that our new EBL is based on, now account better for the coupling
between the supermassive black holes and their host galaxies.
``Bright mode'' AGN feedback may regulate BH formation and temporarily
quench star formation, but it is not a viable long-term mechanism to
keep galaxies red by preventing them from forming stars, as
observations require.  Low-accretion rate ``radio mode'' feedback is a
promising mechanism for counteracting cooling flows and preventing
star formation in clusters and galaxies over long time scales.  Such
SAMs now accurately predict galaxy number counts and luminosity
functions in all spectral bands and all redshifts except for sub-mm
galaxies.  They should therefore allow us to predict the EBL rather
reliably across the wavelengths relevant for gamma-ray attenuation,
since the far-IR EBL is not relevant for attenuation of gamma-rays of
even the highest observed energies.  The predicted range of the EBL is
consistent with the latest estimates of EBL evolution inferred from
observations \cite{franceschini08}.  Thus we have reason to hope that this
sort of theoretical calculation of the EBL and corresponding
gamma-ray attenuation is beginning to converge.

There are two main problems.  One is that the treatment of the crucial
near-IR and mid-IR emission of galaxies by SAMs is still rather
primitive.  The approach that we and others use is based on assuming
that most galaxies of similar total bolometric luminosity emit
radiation with similar spectral energy distributions, so that we can
use ``template'' SEDs to calculate the IR emission.  However, Spitzer
observations have shown that even nearby this is not true, and we have
good reasons to expect that it will be even less true at higher
redshifts.  We are therefore working on self-consistent models of dust
absorption and re-radiation, following on \cite{jonsson06,jonsson06a}
which will be incorporated into future models of the EBL.

The other main problem is calculating the ultraviolet EBL and the
corresponding attenuation of gamma-rays at energies below about 100
GeV, which will be observed by Fermi (GLAST) and by the new generation
of very large ground-based atmospheric Cherenkov telescopes. Little is
known from direct measurements about the EBL at energies above the
Lyman limit. We do not know the relative importance of ionizing
radiation from AGN vs. stars and the redshift evolution of both.
Also, most ionizing photons from star-forming galaxies are absorbed by
local cold gas and dust, with an uncertain fraction escaping to the
intergalactic medium.  The transmission of the IGM is also not fully
understood.  Predicting optical depths to lower energy photons is
further complicated by the fact that the attenuation edge for an
optical depth of unity increases to redshifts of several, meaning that
the evolution of ionizing sources must be understood to high redshift.
Uncertainty in star-formation rates and efficiency, evolution of the
quasar luminosity function and spectrum, and the possibly changing
escape fraction or initial mass function all make predicting optical
depths for gamma rays at high redshift much more difficult than
studies of local absorption at TeV energies. Preliminary calculations
are presented by Gilmore et al. \citep{gilmore08} in this volume, including
predictions for gamma-ray attenuation.

The Fermi satellite is now performing beautifully, frequently
discovering new, or newly brightened, sources of gamma rays.  The
prospect of many new observations in the near future of gamma-rays
across a wide range of energies from sources at many redshifts means
that we can expect a great deal of progress as theory and observations
together clarify the entire spectrum of galactic radiation and its
sources.

\begin{theacknowledgments}
We wish to thank Felix Aharonian, Alberto Dominguez Diaz, Amy Furniss, A. Nepomuk Otte and David A. Williams for helpful discussions concerning gamma--ray attenuation.  Fabio Fontanot assisted in comparisons of different dust templates.  Rudy Gilmore was supported by an AAS International Travel Grant for this conference.  Joel Primack acknowledges support from NASA ATP grant NNX07AG94G, and Rudy Gilmore by NSF-AST-0607712.

\end{theacknowledgments}

\bibliographystyle{aipproc}

\begin{thebibliography}{103}
\expandafter\ifx\csname natexlab\endcsname\relax\def\natexlab#1{#1}\fi
\providecommand{\enquote}[1]{``#1''}
\expandafter\ifx\csname url\endcsname\relax
  \def\url#1{\texttt{#1}}\fi
\expandafter\ifx\csname urlprefix\endcsname\relax\def\urlprefix{URL }\fi
\providecommand{\eprint}[2][]{\url{#2}}

\bibitem[{Hauser} and {Dwek}(2001)]{hauser&dwek01}
M.~G. {Hauser}, and E.~{Dwek}, \emph{\araa} \textbf{39}, 249--307 (2001).

\bibitem[{Bernstein} et~al.(2002{\natexlab{a}})]{bernstein02a}
R.~A. {Bernstein}, W.~L. {Freedman}, and B.~F. {Madore}, \emph{\apj}
  \textbf{571}, 56--84 (2002{\natexlab{a}}).

\bibitem[{Bernstein} et~al.(2002{\natexlab{b}})]{bernstein02b}
R.~A. {Bernstein}, W.~L. {Freedman}, and B.~F. {Madore}, \emph{\apj}
  \textbf{571}, 107--128 (2002{\natexlab{b}}).

\bibitem[{Bernstein}(2007)]{bernstein07}
R.~A. {Bernstein}, \emph{\apj} \textbf{666}, 663--673 (2007).

\bibitem[{Wright} and {Reese}(2000)]{wright&reese00}
E.~L. {Wright}, and E.~D. {Reese}, \emph{\apj} \textbf{545}, 43--55 (2000).

\bibitem[{Wright}(2001)]{wright01}
E.~L. {Wright}, \emph{\apj} \textbf{553}, 538--544 (2001).

\bibitem[{Gorjian} et~al.(2000)]{gorjian00}
V.~{Gorjian}, E.~L. {Wright}, and R.~R. {Chary}, \emph{\apj} \textbf{536},
  550--560 (2000).

\bibitem[{Cambr{\'e}sy} et~al.(2001)]{cambresy01}
L.~{Cambr{\'e}sy}, W.~T. {Reach}, C.~A. {Beichman}, and T.~H. {Jarrett},
  \emph{\apj} \textbf{555}, 563--571 (2001).

\bibitem[{Levenson} et~al.(2007)]{levenson07}
L.~R. {Levenson}, E.~L. {Wright}, and B.~D. {Johnson}, \emph{\apj}
  \textbf{666}, 34--44 (2007).

\bibitem[{Matsumoto} et~al.(2005)]{matsumoto05}
T.~{Matsumoto}, S.~{Matsuura}, H.~{Murakami}, M.~{Tanaka}, M.~{Freund},
  M.~{Lim}, M.~{Cohen}, M.~{Kawada}, and M.~{Noda}, \emph{\apj} \textbf{626},
  31--43 (2005).

\bibitem[{Kashlinsky} and {Odenwald}(2000)]{kashlinsky00}
A.~{Kashlinsky}, and S.~{Odenwald}, \emph{\apj} \textbf{528}, 74--95 (2000).

\bibitem[{Thompson} et~al.(2008)]{thompson08}
R.~I. {Thompson}, D.~{Eisenstein}, X.~{Fan}, M.~{Rieke}, and R.~{Kennicutt},
  (2008), \eprint{astro-ph/0801.3825}.

\bibitem[{Cooray} et~al.(2007)]{cooray07}
A.~{Cooray}, I.~{Sullivan}, R.-R. {Chary}, J.~J. {Bock}, M.~{Dickinson}, H.~C.
  {Ferguson}, B.~{Keating}, A.~{Lange}, and E.~L. {Wright}, \emph{\apjl}
  \textbf{659}, L91--L94 (2007).

\bibitem[{Fixsen} et~al.(1998)]{fixsen98}
D.~J. {Fixsen}, E.~{Dwek}, J.~C. {Mather}, C.~L. {Bennett}, and R.~A. {Shafer},
  \emph{\apj} \textbf{508}, 123--128 (1998).

\bibitem[{Hauser} et~al.(1998)]{hauser98}
M.~G. {Hauser}, R.~G. {Arendt}, T.~{Kelsall}, E.~{Dwek}, N.~{Odegard}, J.~L.
  {Weiland}, H.~T. {Freudenreich}, W.~T. {Reach}, {\it et al.}, \emph{\apj}
  \textbf{508}, 25--43 (1998).

\bibitem[{Finkbeiner} et~al.(2000)]{finkbeiner00}
D.~P. {Finkbeiner}, M.~{Davis}, and D.~J. {Schlegel}, \emph{\apj} \textbf{544},
  81--97 (2000).

\bibitem[{Lagache} et~al.(2000)]{lagache00}
G.~{Lagache}, L.~M. {Haffner}, R.~J. {Reynolds}, and S.~L. {Tufte}, \emph{\aap}
  \textbf{354}, 247--252 (2000).

\bibitem[{Wright}(2004)]{wright04}
E.~L. {Wright}, \emph{New Astronomy Review} \textbf{48}, 465--468 (2004).

\bibitem[{Puget} and {Lagache}(2001)]{puget&lagache01}
J.~L. {Puget}, and G.~{Lagache}, \enquote{{Far-infrared Source Counts and the
  Diffuse Infrared Background},} in \emph{The Extragalactic Infrared Background
  and its Cosmological Implications}, edited by M.~{Harwit}, 2001, vol. 204 of
  \emph{IAU Symposium}, pp. 233.

\bibitem[{Madau} and {Pozzetti}(2000)]{madau00}
P.~{Madau}, and L.~{Pozzetti}, \emph{\mnras} \textbf{312}, L9--L15 (2000).

\bibitem[{Totani} et~al.(2001)]{totani01}
T.~{Totani}, Y.~{Yoshii}, F.~{Iwamuro}, T.~{Maihara}, and K.~{Motohara},
  \emph{\apjl} \textbf{550}, L137--L141 (2001).

\bibitem[{Xu} et~al.(2005)]{xu05}
C.~K. {Xu}, J.~{Donas}, S.~{Arnouts}, T.~K. {Wyder}, M.~{Seibert},
  J.~{Iglesias-P{\'a}ramo}, J.~{Blaizot}, T.~{Small}, {\it et al.}, \emph{\apjl}
  \textbf{619}, L11--L14 (2005).

\bibitem[{Gardner} et~al.(2000)]{gardner00}
J.~P. {Gardner}, T.~M. {Brown}, and H.~C. {Ferguson}, \emph{\apjl}
  \textbf{542}, L79--L82 (2000).

\bibitem[{Fazio} et~al.(2004)]{fazio04}
G.~G. {Fazio}, M.~L.~N. {Ashby}, P.~{Barmby}, J.~L. {Hora}, J.-S. {Huang},
  M.~A. {Pahre}, Z.~{Wang}, S.~P. {Willner}, {\it et al.}, \emph{\apjs} \textbf{154}, 39--43 (2004).

\bibitem[{Levenson} and {Wright}(2008)]{levenson08}
L.~R. {Levenson}, and E.~L. {Wright}, (2008),
  \eprint{astro-ph/0802.1239}.

\bibitem[{Elbaz} et~al.(2002)]{elbaz02}
D.~{Elbaz}, C.~J. {Cesarsky}, P.~{Chanial}, H.~{Aussel}, A.~{Franceschini},
  D.~{Fadda}, and R.~R. {Chary}, \emph{\aap} \textbf{384}, 848--865 (2002).

\bibitem[{Chary} et~al.(2004)]{chary04}
R.~{Chary}, S.~{Casertano}, M.~E. {Dickinson}, H.~C. {Ferguson}, P.~R.~M.
  {Eisenhardt}, D.~{Elbaz}, N.~A. {Grogin}, L.~A. {Moustakas}, W.~T. {Reach},
  and H.~{Yan}, \emph{\apjs} \textbf{154}, 80--86 (2004).

\bibitem[{Frayer} et~al.(2006)]{frayer06}
D.~T. {Frayer}, M.~T. {Huynh}, R.~{Chary}, M.~{Dickinson}, D.~{Elbaz},
  D.~{Fadda}, J.~A. {Surace}, H.~I. {Teplitz}, L.~{Yan}, and B.~{Mobasher},
  \emph{\apjl} \textbf{647}, L9--L12 (2006).

\bibitem[{Papovich} et~al.(2004)]{papovich04}
C.~{Papovich}, H.~{Dole}, E.~{Egami}, E.~{Le Floc'h}, P.~G.
  {P{\'e}rez-Gonz{\'a}lez}, A.~{Alonso-Herrero}, L.~{Bai}, C.~A. {Beichman},
 {\it et al.}, \emph{\apjs} \textbf{154}, 70--74 (2004).

\bibitem[{Dole} et~al.(2006)]{dole06}
H.~{Dole}, G.~{Lagache}, J.-L. {Puget}, K.~I. {Caputi},
  N.~{Fern{\'a}ndez-Conde}, E.~{Le Floc'h}, C.~{Papovich}, P.~G., {\it et al.}, \emph{\aap}
  \textbf{451}, 417--429 (2006).

\bibitem[{Coppin} et~al.(2006)]{coppin06}
K.~{Coppin}, E.~L. {Chapin}, A.~M.~J. {Mortier}, S.~E. {Scott}, C.~{Borys},
  J.~S. {Dunlop}, M.~{Halpern}, D.~H. {Hughes}, {\it et al.},
  \emph{\mnras} \textbf{372}, 1621--1652 (2006).

\bibitem[{Primack} et~al.(2005)]{primack05}
J.~R. {Primack}, J.~S. {Bullock}, and R.~S. {Somerville},
  \enquote{{Observational Gamma-ray Cosmology},} in \emph{High Energy Gamma-Ray
  Astronomy}, edited by F.~A. {Aharonian}, H.~J. {V{\"o}lk}, and D.~{Horns},
  2005, vol. 745 of \emph{American Institute of Physics Conference Series}, pp.
  23--33.

\bibitem[{Franceschini} et~al.(2008)]{franceschini08}
A.~{Franceschini}, G.~{Rodighiero}, and M.~{Vaccari}, (2008), \eprint{astro-ph/0805.1841}.

\bibitem[{Dolch}(2008)]{dolch08}
T.~{Dolch}, \emph{in prep}  (2008).

\bibitem[{Jones} et~al.(2006)]{jones06}
D.~H. {Jones}, B.~A. {Peterson}, M.~{Colless}, and W.~{Saunders}, \emph{\mnras}
  \textbf{369}, 25--42 (2006).

\bibitem[{Cole} et~al.(2001)]{cole01}
S.~{Cole}, P.~{Norberg}, C.~M. {Baugh}, C.~S. {Frenk}, J.~{Bland-Hawthorn},
  T.~{Bridges}, R.~{Cannon}, M.~{Colless}, {\it et al.}, \emph{\mnras} \textbf{326}, 255--273
  (2001).

\bibitem[{Bell} et~al.(2003)]{bell03}
E.~F. {Bell}, D.~H. {McIntosh}, N.~{Katz}, and M.~D. {Weinberg}, \emph{\apjs}
  \textbf{149}, 289--312 (2003).

\bibitem[{Takeuchi} et~al.(2001)]{takeuchi01}
T.~T. {Takeuchi}, T.~T. {Ishii}, H.~{Hirashita}, K.~{Yoshikawa},
  H.~{Matsuhara}, K.~{Kawara}, and H.~{Okuda}, \emph{\pasj} \textbf{53}, 37--52
  (2001).

\bibitem[{Hopkins}(2004)]{hopkins04}
A.~M. {Hopkins}, \emph{\apj} \textbf{615}, 209--221 (2004).

\bibitem[{Gabasch} et~al.(2004)]{gabasch04}
A.~{Gabasch}, M.~{Salvato}, R.~P. {Saglia}, R.~{Bender}, U.~{Hopp}, S.~{Seitz},
  G.~{Feulner}, M.~{Pannella}, {\it et al.},
  \emph{\apjl} \textbf{616}, L83--L86 (2004).

\bibitem[{Hopkins} and {Beacom}(2006)]{hopkins&beacom06}
A.~M. {Hopkins}, and J.~F. {Beacom}, \emph{\apj} \textbf{651}, 142--154 (2006).

\bibitem[{Fardal} et~al.(2007)]{fardal07}
M.~A. {Fardal}, N.~{Katz}, D.~H. {Weinberg}, and R.~{Dav{\'e}}, \emph{\mnras}
  \textbf{379}, 985--1002 (2007).

\bibitem[{Dav{\'e}}(2008)]{dave08}
R.~{Dav{\'e}}, \emph{\mnras} \textbf{385}, 147--160 (2008).

\bibitem[{Chen} et~al.(2008)]{chen08}
Y.-M. {Chen}, V.~{Wild}, G.~{Kauffmann}, J.~{Blaizot}, M.~{Davis}, K.~{Noeske},
  J.-M. {Wang}, and C.~{Willmer}, (2008),
  \eprint{astro-ph/0808.3683}.

\bibitem[{Salvaterra} and {Ferrara}(2003)]{salvaterra&ferrara03}
R.~{Salvaterra}, and A.~{Ferrara}, \emph{\mnras} \textbf{339}, 973--982 (2003).

\bibitem[{Salvaterra} and {Ferrara}(2006)]{salvaterra&ferrara06}
R.~{Salvaterra}, and A.~{Ferrara}, \emph{\mnras} \textbf{367}, L11--L15 (2006).

\bibitem[{Dwek} et~al.(2005)]{dwek05}
E.~{Dwek}, R.~G. {Arendt}, and F.~{Krennrich}, \emph{\apj} \textbf{635},
  784--794 (2005).

\bibitem[{Madau} and {Silk}(2005)]{madau&silk05}
P.~{Madau}, and J.~{Silk}, \emph{\mnras} \textbf{359}, L37--L41 (2005).

\bibitem[{Kneiske} et~al.(2002)]{kneiske02}
T.~M. {Kneiske}, K.~{Mannheim}, and D.~H. {Hartmann}, \emph{\aap} \textbf{386},
  1--11 (2002).

\bibitem[{Stecker} et~al.(2006)]{steckermalkan&scully06}
F.~W. {Stecker}, M.~A. {Malkan}, and S.~T. {Scully}, \emph{\apj} \textbf{648},
  774--783 (2006).

\bibitem[{Primack} et~al.(2001)]{primack01}
J.~R. {Primack}, R.~S. {Somerville}, J.~S. {Bullock}, and J.~E.~G. {Devriendt},
  \enquote{{Probing Galaxy Formation with High-Energy Gamma Rays},} in
  \emph{American Institute of Physics Conference Series}, edited by F.~A.
  {Aharonian}, and H.~J. {V{\"o}lk}, 2001, vol. 558 of \emph{American Institute
  of Physics Conference Series}, pp. 463.

\bibitem[{Somerville} et~al.(2008)]{somerville08}
R.~S. {Somerville}, P.~F. {Hopkins}, T.~J. {Cox}, B.~E. {Robertson}, and
  L.~{Hernquist}, (2008), \eprint{astro-ph/0808.1227}.

\bibitem[{Schiminovich} et~al.(2005)]{schiminovich05}
D.~{Schiminovich}, O.~{Ilbert}, S.~{Arnouts}, B.~{Milliard}, L.~{Tresse},
  O.~{Le F{\`e}vre}, M.~{Treyer}, T.~K. {Wyder},
  {\it et al.}, \emph{\apjl} \textbf{619},
  L47--L50 (2005).

\bibitem[{Fontana} et~al.(2006)]{fontana06}
A.~{Fontana}, S.~{Salimbeni}, A.~{Grazian}, E.~{Giallongo}, L.~{Pentericci},
  M.~{Nonino}, F.~{Fontanot}, N.~{Menci}, {\it et al.}, \emph{\aap} \textbf{459},
  745--757 (2006).

\bibitem[{Borch} et~al.(2006)]{borch06}
A.~{Borch}, K.~{Meisenheimer}, E.~F. {Bell}, H.-W. {Rix}, C.~{Wolf}, S.~{Dye},
  M.~{Kleinheinrich}, Z.~{Kovacs}, and L.~{Wisotzki}, \emph{\aap} \textbf{453},
  869--881 (2006).

\bibitem[{Wilkins} et~al.(2008)]{wilkins08}
S.~M. {Wilkins}, N.~{Trentham}, and A.~M. {Hopkins}, \emph{\mnras}
  \textbf{385}, 687--694 (2008).

\bibitem[{Montero-Dorta} and {Prada}(2008)]{md&prada08}
A.~D. {Montero-Dorta}, and F.~{Prada}, (2008),
  \eprint{astro-ph/0806.4930}.

\bibitem[{Sanders} et~al.(2007)]{sanders07}
D.~B. {Sanders}, M.~{Salvato}, H.~{Aussel}, O.~{Ilbert}, N.~{Scoville}, J.~A.
  {Surace}, D.~T. {Frayer}, K.~{Sheth}, {\it et al.}, \emph{\apjs} \textbf{172}, 86--98 (2007).

\bibitem[{Shupe} et~al.(2008)]{shupe08}
D.~L. {Shupe}, M.~{Rowan-Robinson}, C.~J. {Lonsdale}, F.~{Masci}, T.~{Evans},
  {\it et al.}, \emph{\aj} \textbf{135}, 1050--1056 (2008).

\bibitem[{White} and {Frenk}(1991)]{white91}
S.~D.~M. {White}, and C.~S. {Frenk}, \emph{\apj} \textbf{379}, 52--79 (1991).

\bibitem[{Somerville} and {Primack}(1999)]{somerville&primack99}
R.~S. {Somerville}, and J.~R. {Primack}, \emph{\mnras} \textbf{310}, 1087--1110
  (1999).

\bibitem[{Somerville} et~al.(2001)]{somerville01}
R.~S. {Somerville}, J.~R. {Primack}, and S.~M. {Faber}, \emph{\mnras}
  \textbf{320}, 504--528 (2001).

\bibitem[{Devriendt} et~al.(1999)]{devriendt99}
J.~E.~G. {Devriendt}, B.~{Guiderdoni}, and R.~{Sadat}, \emph{\aap}
  \textbf{350}, 381--398 (1999).

\bibitem[{Devriendt} and {Guiderdoni}(2000)]{devriendt00}
J.~E.~G. {Devriendt}, and B.~{Guiderdoni}, \emph{\aap} \textbf{363}, 851--862
  (2000).

\bibitem[{Blanton} et~al.(2003)]{blanton03}
M.~R. {Blanton}, D.~W. {Hogg}, N.~A. {Bahcall}, J.~{Brinkmann}, M.~{Britton},
  A.~J. {Connolly}, I.~{Csabai}, M.~{Fukugita}, {\it et al.}, \emph{\apj} \textbf{592}, 819--838 (2003).

\bibitem[{Norberg} et~al.(2002)]{norberg02}
P.~{Norberg}, C.~M. {Baugh}, E.~{Hawkins}, S.~{Maddox}, D.~{Madgwick},
  O.~{Lahav}, S.~{Cole}, C.~S. {Frenk}, {\it et al.}, \emph{\mnras}
  \textbf{332}, 827--838 (2002).

\bibitem[{Komatsu} et~al.(2008)]{komatsu08}
E.~{Komatsu}, J.~{Dunkley}, M.~R. {Nolta}, C.~L. {Bennett}, B.~{Gold},
  G.~{Hinshaw}, N.~{Jarosik}, D.~{Larson}, {\it et al.}, (2008), \eprint{astro-ph/0803.0547}.

\bibitem[{Somerville} and {Kolatt}(1999)]{somerville&kolatt99}
R.~S. {Somerville}, and T.~S. {Kolatt}, \emph{\mnras} \textbf{305}, 1--14
  (1999).

\bibitem[{Navarro} et~al.(1997)]{navarro97}
J.~F. {Navarro}, C.~S. {Frenk}, and S.~D.~M. {White}, \emph{\apj} \textbf{490},
  493 (1997).

\bibitem[{Bullock} et~al.(2001)]{bullock01}
J.~S. {Bullock}, T.~S. {Kolatt}, Y.~{Sigad}, R.~S. {Somerville}, A.~V.
  {Kravtsov}, A.~A. {Klypin}, J.~R. {Primack}, and A.~{Dekel}, \emph{\mnras}
  \textbf{321}, 559--575 (2001).

\bibitem[{Birnboim} and {Dekel}(2003)]{birnboim&dekel03}
Y.~{Birnboim}, and A.~{Dekel}, \emph{\mnras} \textbf{345}, 349--364 (2003).

\bibitem[{Dekel} et~al.(2008)]{dekel&birnboim08}
A.~{Dekel}, Y.~{Birnboim}, G.~{Engel}, J.~{Freundlich}, T.~{Goerdt},
  M.~{Mumcuoglu}, E.~{Neistein}, C.~{Pichon}, R.~{Teyssier}, and E.~{Zinger},
 (2008), \eprint{astro-ph/0808.0553}.

\bibitem[{Keres} et~al.(2008)]{keres08}
D.~{Keres}, N.~{Katz}, M.~{Fardal}, R.~{Dave}, and D.~H. {Weinberg},
 (2008), \eprint{astro-ph/0809.1430}.

\bibitem[{Kennicutt}(1989)]{kennicutt89}
R.~C. {Kennicutt}, Jr., \emph{\apj} \textbf{344}, 685--703 (1989).

\bibitem[{Kennicutt} et~al.(1998)]{kennicutt98}
R.~C. {Kennicutt}, Jr., P.~B. {Stetson}, A.~{Saha}, D.~{Kelson}, D.~M.
  {Rawson}, S.~{Sakai}, B.~F. {Madore}, J.~R. {Mould},   {\it et al.}, \emph{\apj} \textbf{498}, 181 (1998).

\bibitem[{Cox} et~al.(2008)]{cox08}
T.~J. {Cox}, P.~{Jonsson}, R.~S. {Somerville}, J.~R. {Primack}, and A.~{Dekel},
  \emph{\mnras} \textbf{384}, 386--409 (2008).

\bibitem[{Sijacki} et~al.(2007)]{sijacki07}
D.~{Sijacki}, V.~{Springel}, T.~{di Matteo}, and L.~{Hernquist}, \emph{\mnras}
  \textbf{380}, 877--900 (2007).

\bibitem[{Bondi}(1952)]{bondi52}
H.~{Bondi}, \emph{\mnras} \textbf{112}, 195 (1952).

\bibitem[{Lagache} et~al.(2005)]{lagache05}
G.~{Lagache}, J.-L. {Puget}, and H.~{Dole}, \emph{\araa} \textbf{43}, 727--768
  (2005).

\bibitem[{Guiderdoni} and {Devriendt}(1999)]{guiderdoni99}
B.~{Guiderdoni}, and J.~E.~G. {Devriendt}, (1999), \eprint{astro-ph/9911162}.

\bibitem[{Gould} and {Schreder}(1967)]{gould67}
R.~J. {Gould}, and G.~P. {Schreder}, \emph{\physrev} \textbf{155}, 1404 (1967).

\bibitem[{Maier}(2007)]{maier07}
G.~{Maier}, (2007), \eprint{astro-ph/0709.3654}.

\bibitem[{Hinton}(2004)]{hinton04}
J.~A. {Hinton}, \emph{New Astronomy Review} \textbf{48}, 331--337 (2004).

\bibitem[{Cortina}(2005)]{cortina05}
J.~{Cortina}, \emph{\apss} \textbf{297}, 245--255 (2005).

\bibitem[{Tavani} et~al.(2008)]{tavani08}
M.~{Tavani}, G.~{Barbiellini}, A.~{Argan}, A.~{Bulgarelli}, P.~{Caraveo},
  A.~{Chen}, V.~{Cocco}, E.~{Costa}, {\it et al.}, \emph{Nuclear
  Instruments and Methods in Physics Research A} \textbf{588}, 52--62 (2008).

\bibitem[{Aharonian} et~al.(2006)]{aharonian06}
F.~{Aharonian}, A.~G. {Akhperjanian}, A.~R. {Bazer-Bachi}, M.~{Beilicke},
  {\it et al.}, \emph{\nat} \textbf{440}, 1018--1021 (2006).

\bibitem[{MAGIC Collaboration} et~al.(2008)]{albert08}
{MAGIC Collaboration}, J.~{Albert}, E.~{Aliu}, H.~{Anderhub}, L.~A.
  {Antonelli}, P.~{Antoranz}, M.~{Backes}, C.~{Baixeras}, J.~A. {Barrio},
  {\it et al.}, \emph{Science} \textbf{320}, 1752 (2008).

\bibitem[{Mazin} and {Raue}(2007)]{mazin&raue07}
D.~{Mazin}, and M.~{Raue}, \emph{\aap} \textbf{471}, 439--452 (2007).

\bibitem[{Dwek} and {Krennrich}(2005)]{dwek&krennrich05}
E.~{Dwek}, and F.~{Krennrich}, \emph{\apj} \textbf{618}, 657--674 (2005).

\bibitem[{Begelman} et~al.(2008)]{begelman08}
M.~C. {Begelman}, A.~C. {Fabian}, and M.~J. {Rees}, \emph{\mnras} \textbf{384},
  L19--L23 (2008).

\bibitem[{Fossati} et~al.(1998)]{fossati98}
G.~{Fossati}, L.~{Maraschi}, A.~{Celotti}, A.~{Comastri}, and G.~{Ghisellini},
  \emph{\mnras} \textbf{299}, 433--448 (1998).

\bibitem[{Ghisellini} et~al.(1998)]{ghisellini98}
G.~{Ghisellini}, A.~{Celotti}, G.~{Fossati}, L.~{Maraschi}, and A.~{Comastri},
  \emph{\mnras} \textbf{301}, 451--468 (1998).

\bibitem[{Maraschi} et~al.(2008)]{maraschi08}
L.~{Maraschi}, G.~{Ghisellini}, F.~{Tavecchio}, L.~{Foschini}, and R.~M.
  {Sambruna}, (2008), \eprint{astro-ph/0802.1789}.

\bibitem[{Ghisellini} et~al.(2002)]{ghisellini02}
G.~{Ghisellini}, A.~{Celotti}, and L.~{Costamante}, \emph{\aap} \textbf{386},
  833--842 (2002).

\bibitem[{Aharonian}(2001)]{aharonian01}
F.~A. {Aharonian}, \enquote{{TeV blazars and cosmic infrared background
  radiation},} in \emph{International Cosmic Ray Conference}, edited by O.~G.
  {Gladysheva}, G.~E. {Kocharov}, G.~A. {Kovaltsov}, and I.~G. {Usoskin}, 2001,
  vol.~27-I of \emph{International Cosmic Ray Conference}, pp. 250.

\bibitem[{Katarzy{\'n}ski} et~al.(2006)]{katarzynski06}
K.~{Katarzy{\'n}ski}, G.~{Ghisellini}, F.~{Tavecchio}, J.~{Gracia}, and
  L.~{Maraschi}, \emph{\mnras} \textbf{368}, L52--L56 (2006).

\bibitem[{Aharonian} et~al.(2007)]{aharonian07}
F.~{Aharonian}, A.~G. {Akhperjanian}, U.~{Barres de Almeida}, A.~R.
  {Bazer-Bachi}, B.~{Behera}, M.~{Beilicke}, W.~{Benbow}, K.~{Bernl{\"o}hr},
  {\it et al.}, \emph{\aap} \textbf{475}, L9--L13
  (2007).

\bibitem[{Teshima} et~al.(2007)]{teshima07}
M.~{Teshima}, E.~{Prandini}, R.~{Bock}, M.~{Errando}, D.~{Kranich},
  {\it et al.}, (2007),
  \eprint{astro-ph/0709.1475}.

\bibitem[{Stecker} and {Scully}(2008)]{stecker08}
F.~W. {Stecker}, and S.~T. {Scully}, (2008),
  \eprint{astro-ph/0807.4880}.

\bibitem[{Stecker} et~al.(2007)]{steckermalkan&scully07}
F.~W. {Stecker}, M.~A. {Malkan}, and S.~T. {Scully}, \emph{\apj} \textbf{658},
  1392--1392 (2007).

\bibitem[{Costamante} et~al.(2004)]{costamante03}
L.~{Costamante}, F.~{Aharonian}, D.~{Horns}, and G.~{Ghisellini}, \emph{New
  Astronomy Review} \textbf{48}, 469--472 (2004).

\bibitem[{Jonsson}(2006)]{jonsson06}
P.~{Jonsson}, \emph{\mnras} \textbf{372}, 2--20 (2006).

\bibitem[{Jonsson} et~al.(2006)]{jonsson06a}
P.~{Jonsson}, T.~J. {Cox}, J.~R. {Primack}, and R.~S. {Somerville}, \emph{\apj}
  \textbf{637}, 255--268 (2006).

\bibitem[{Gilmore} et~al.(2008)]{gilmore08}
R.~C. {Gilmore}, P. {Madau}, J.~R. {Primack}, and R.~S. {Somerville}, \enquote{{Modeling Gamma-Ray Attenuation in High Redshift GeV Spectra},} to be published in these proceedings,
  \eprint{astro-ph/0811.1984}.

\end{thebibliography}

\end{document}